# IOTA Feasibility and Perspectives for Enabling Vehicular Applications

Paulo C. Bartolomeu[1,2], Emanuel Vieira[2], and Joaquim Ferreira[1,3]

[1]Instituto de Telecomunicações, 3810-193 Aveiro, Portugal
[2]DETI - University of Aveiro, 3810-193 Aveiro, Portugal
[3]ESTGA - University of Aveiro, 3750-127 Águeda, Portugal

*Abstract*—The emergence of distributed ledger technologies in the vehicular applications' arena is decisively contributing to their improvement and shaping of the public opinion about their future. The Tangle is a technology at its infancy, but showing enormous potential to become a key solution by addressing several of the blockchain's limitations. This paper focuses on the use of the Tangle to improve the security of both in-vehicle and off-vehicle functions in vehicular applications. To this end, key operational performance parameters are identified, evaluated and discussed with emphasis on their limitations and potential impact in future vehicular applications.

## I. INTRODUCTION

The last decade has been fertile for vehicular research. Electrical vehicles became a reality and a trend due to the tightening of pollutant emission legislation. Vehicles became autonomous and (wireless) communications have evolved to provide connectivity with quality of service and flexibility tailored to enable such applications (e.g., through 5G slicing). Today it is not difficult to conceive a world where families do not own cars, but are transported by them in an intelligent fashion. A world where car fleets manage themselves according to the foreseen usage and where cars transport people from their homes to work so as to avoid all possible energy waste and maximize profitability of their owners while providing maximum comfort and added experiences to its users.

The future of transport and mobility will surely be enabled by autonomous vehicles encompassing sensing capabilities that can cooperate with each other and share their sensing resources and perspective with the nearby infrastructures and neighboring vehicles. However, many challenges remain ahead regarding key aspects such as security, privacy and anonymity. Anonymity, as the ability to share information without disclosing the identity of its producer, is particularly relevant when dealing with the location of vehicles, for example, where pseudonyms are typically used to avoid exposing the identity of the vehicle that would otherwise allow collecting information about its location or traveled routes. Privacy deals with the ability to keep the data private, i.e., of ensuring that even if someone gets access to the raw data he/she cannot make sense of it. For example, communications among people in different vehicles via the infotainment system must be conducted so as to guarantee that only the authorized people are able to listen and participate in the conversations. Finally, security is the ability to avoid threats that can compromise the system. A secure system ensures that user access rules are enforced and that mechanisms are in place to avoid attacks from endanger the security of people (e.g., passengers in vehicles and/or pedestrians on the road).

The security of vehicles can be classified in two operational domains: in-vehicle and off-vehicle. The first case encompasses functions that are intrinsically running inside the vehicle. Examples of safety critical functions in this context are steering, breaking, localization and environment sensing. These can be used to support autonomous driving and, therefore, must fulfill stringent dependability and security requirements. Other functions such as infotainment, acclimatization, air quality have less stringent requirements because, although having the potential to negatively impact the user experience on-board, they are less likely to endanger human lives. These are considered functions with mission critical security requirements. The second case, off-vehicle security, encompasses functions that are externally made available to the vehicle and that can be used to complement its own in-vehicle functions. A typical example of such a function is the cooperative sensing and perception shared among vehicles using CAMs and DENMs. Off-vehicle functions have typically mission critical security requirements.

In the last years, there has been numerous reports of hacking of in-vehicle systems. In 2015, Charlie Miller

This work is supported by the European Regional Development Fund (FEDER), through the Regional Operational Programme of Centre (CENTRO 2020) of the Portugal 2020 framework [Project PASMO with Nr. 000008 (CENTRO-01-0246-FEDER-000008)].
Corresponding author: Paulo C. Bartolomeu (bartolomeu@ua.pt)

and Chris Valasek have demonstrated that they could take control over the Internet of a Jeep Cherokee's entertainment system by meddling with the dashboard functions, steering, brakes, and transmission, all from a laptop that could be anywhere in the world [3]. In 2016, researchers from Chinas Zhejiang University and from a Chinese security firm named Qihoo 360 have proved that it is possible to jam multiple sensors on a Tesla Model S making objects invisible to its navigation system [1], something that can severely jeopardize its autonomous navigation functions. Recently, in April this year, Daan Keuper and Thijs Alkemad revealed that it is possible to hack a Volkswagen Golf GTE and an Audi A3 Sportback e-tron via a Wi-Fi connection [4]. In their demonstration, they gained access to the In-Vehicle Infotainment system, which allow them to listen to conversations in the vehicle through the car kit, access the address book and the conversation history, besides tracking the location of the vehicle and past history based on the navigation system. More recently in May, 14 vulnerabilities have been discovered in BMW vehicles, 6 of them providing remote access to the car via the wireless Bluetooth and cellular network interfaces [2].

Vehicular applications have been pushing research efforts and standardization initiatives leading to the arise of several security protocols and Public Key Infrastructure (PKI) architectures. The most relevant ones are the IEEE Wireless Access in Vehicular Environments (WAVE) 1609.2 [6] (USA) and ETSI ITS Security Standards (Europe) [7]. Due to the relative early status of adoption of these standards and slow emergence of cooperative sensing applications, no accounts of security breaches have been yet reported to the best of our knowledge. Nevertheless, efforts are being conducted in this domain, both in terms of characterizing the performance of standard security mechanisms [8] as well as proposing new security solutions that complement existing standards [9].

Over the last few years the blockchain technology has managed to attract immense attention due to intrinsic properties such as trustless operation, immutability, transparency, easy verification, cryptographic security, auditability and independence of third parties [10]. Blockchain is perceived has having the potential to perform a radical change in several sectors, most notably in those where it can harness synergies with emerging technologies such as machine learning, artificial intelligence, autonomous driving and fog computing [11]. However, the adoption of public blockchain technologies presents several drawbacks that must be carefully considered [5].

This paper addresses the use of the IOTA Distributed Ledger Technology (DLT) [20] to cope with the shortcomings of existing public blockchain technologies, focusing on its timeliness to improve the security of both in-vehicle and off-vehicle functions. Section II provides an introduction to distributed ledger technologies with an emphasis on existing public blockchains and IOTA. In the first case, an overview of representative blockchain applications is provided. In the second case, an introduction to the IOTA cryptocurrency is documented. Section III describes the setup used to collect relevant performance data about the *Tangle*, the results and their discussion. Finally, Section IV concludes this paper and provides future work directions.

## II. DISTRIBUTED LEDGER TECHNOLOGIES

In the last few years we have been assisting a rise in the adoption of Distributed Ledger Technologies (DLTs) in the vehicular domain. In this section an overview of public blockchain applications to vehicular communications is provided. Furthermore, as a promising DLT that aims at solving the blockchain's shortcomings, the *Tangle* operation is introduced.

### A. Public Blockchain Applications

A new cryptocurrency named Bitcoin was introduced in 2008 [12] encompassing a mechanism named blockchain. The key innovation introduced by the blockchain was the elimination of the central authority that was responsible for guaranteeing secure and valid exchanges while being liable when that did not occur. A public decentralized peer-to-peer ledger system was introduced where all transactions reside in a distributed database (blockchain data structure) and are validated by all peers in the network via a consensus protocol. A blockchain can be viewed both as an information and communications technology aimed at managing the ownership of assets and of the related rights/obligations and as a mechanism to "to decentralize the governance structures used to coordinate people and economic decision making" [11].

The vehicular ecosystem offers a broad diversity of opportunities to capitalize on the benefits provided by blockchain technologies. Regarding certificate management, existing solutions to secure inter-vehicle communications mainly rely on the use of digital certificates for authentication, whose validation must be performed within a strict time bound. Not only this approach embodies stringent requirements in terms of computation power in all nodes, it also introduces a single point of failure in the central node that issues and revokes certificates. To cope with these limitations, several proposals have been put forward to manage certificates [13] and keys [15] in distributed and immutable records, and supervise remote software updates or provide dynamic insurance fees [14].

Another application domain that is becoming popular for adopting blockchain technologies is the management of trust in a vehicular networks. This domain stemmed from the idea that the data's credibility can be estimated by analyzing the past behavior of its producers. In one example the blockchain system is used to rate vehicles that participate in a network by analyzing the received message sensing content, performing its perceptual validation and globally establishing the credibility of its issuer [16]. A mechanism addressing the same goal has been proposed by [17]. In this case, a trust parameter named *Trust Bit* is employed to provide information regarding the messages' level of trust as perceived by the vehicle network.

One major issue in vehicular networks is the lack of "motivation" to forward announcements among vehicles. This occurs due to the absence of incentives to compensate the energy expenditure and computational resources involved in the operation. To overcome this issue two solutions have been proposed in [18] and [19]. These solutions stimulate the cooperation of vehicles for the task of relaying messages.

As discussed, an increasing number of vehicular applications are adopting public blockchain technologies. However, their use brings multiple concerns that must be carefully considered, namely their reliance on a private key, immutability, bias towards nodes (or sets of nodes) with very high computing power, high latency and lack of intrinsic societal value for executing "proof-of-work" [5]. The *Tangle* has been created to overcome these shortcomings and it is presented next.

### B. The Tangle

The IOTA is a cryptocurrency that was created with a focus on the Internet of Things to solve the problems of scalability, control centralization and post-quantum security that characterizes other cryptocurrencies employing the blockchain technology. The *Tangle* is its key contribution and builds on the concept of Directed Acyclic Graphs (DAGs) to substitute to blockchains.

In the *Tangle* each node is a transaction. For a transaction to be added to the *Tangle* it must approve two other transactions by doing a small amount of "proof-of-work". All transactions which have not been yet approved are called tips. A Markov Chain Monte Carlo algorithm is used to chose the tips that will be submitted for approval in the *Tangle* [20]. Starting from a random transaction *A* the algorithm selects another transaction which approves *A* with a bias towards the ones with larger cumulative weight. The algorithm keeps running until it reaches a tip. Running through transactions with larger cumulative weight ensures that the most *worthy* tips (i.e. the tips with a larger PoW) are chosen to be approved.

Currently, the *Tangle* is small in size. Hence, entities with malicious intents and sufficient computational power when compared to other users can carry out attacks such as double-spending transactions. In this type of attack, the attacker creates two different outgoing transactions in the same timeframe in a effort to spend his total wallet balance two times, effectively becoming negative. Using sufficient computational power the attacker could create a considerable amount of transactions that would directly and indirectly approve the double-spending ones thus rendering them reputable and accepted by the *Tangle* network. Due to the current small size of the *Tangle* anyone with sufficient computational power can carry out attacks such as these. Therefore, due to this *Tangle* "Beta" status, the exact rules of tip selection are not publicly available currently. A *central* node called *Coordinator* is given the role of electing the tips to approve while the *Tangle* scale is not large enough to guarantee its independent operation.

A transaction approved by the *Coordinator* is said to be "Confirmed". The process of electing tips to approve occurs with a period of one minute and the result is a transaction called "Milestone". This transaction is similar to any other transaction, i.e., it approves two other transactions, but contains the *Coordinator's* signature. A particular transaction is deemed "Confirmed" if the latest Milestone enforces its approval, directly or indirectly (there exists an approval path leading to the latest Milestone).

Every transaction can carry a message. This allows for two parties to communicate between each other. The drawback is, since anyone can see every transaction in the *Tangle* this type of communication is not really feasible when the content must be kept confidential. Developed by the IOTA Foundation, a messaging protocol was introduced called Masked Authenticated Messaging (MAM). This protocol uses the *Tangle* network as one normally does, communicating by adding transactions, but with an extra layer of encryption. Messages are encrypted before adding the transactions on to the *Tangle*. In this case only the target party (or parties) can decode the encrypted message of the appended transaction.

The next section describes the setup used to conduct an evaluation of several *Tangle* key performance parameters and documents the obtained results and discussion.

### III. EVALUATION

This section globally describes the used setup and the preliminary results of the time required to append transactions to the *Tangle*. This will provide the grounds to discuss the feasibility of implementing vehicular applications with IOTA.

*A. Setup*

In order to evaluate the transaction writing times in the *Tangle* a basic test setup encompassing two nodes was used: a public node at *https://wallet2.iota.town:443* (Node A), hosted in Norway, and a private node (node B) hosted in a Virtual Private Server (VPS) in Germany. The latter was connected to the *Tangle* network using the *CarrIOTA Nelson* project.

Regarding the hardware, the private node employs a 4 core Xeon CPU E5-2620 v3 running at 2.40GHz, with a 12 GB RAM memory and a 300 GB SSD disk for storage. The hardware specification of the public node was not available at the time of writing of this paper.

Using the IOTA's Python API, PyOTA, transactions with payloads of different lengths were added in order to assess their influence on the transaction writing time. An IOTA transaction is characterized by a payload length of 2187 trytes. If the content to be stored into the transaction is larger than 2187 trytes, it is segmented in two transactions added to the *Tangle*. In our case two message lengths were considered: $u$ with 1093 trytes and $m$ with 2405 trytes.

Three different scenarios have been considered regarding the Tangle performance: appending transactions and appending Masked Authenticated Messages. Their specific testing conditions are detailed bellow:

*1) Append transactions to the Tangle:* The tests were conducted in two phases. In the first, a set of 100 trials was conducted and the overal delay of adding these transactions to the tangle was measured. In the second phase, also for a set of 100 trials, in order to better understand the delays experienced at each stage of the transaction creation, the delay measurement was segmented in terms of its "tip selection", "attach to tangle" and "broadcast" operations.

*2) Append MAMs to the Tangle:* The tests were conducted for a set of 100 trials by measuring the overall delay of appending a Masked Authenticated Message to the *Tangle* segmented in terms of its "encoding", "tip selection", "attach to tangle","broadcast" and "get message" operations.

*B. Results*

The collected results are organized according to the previous identified scenarios: appending transactions and appending Masked Authenticated Messages.

*1) Append transactions to the Tangle:* The box-and-whisker plot of the global delay for attaching a transaction to the *Tangle* is documented in Fig. 1. As expected, for both nodes (A and B), the message with the highest length ($m$) is characterized by an higher median delay to be added to the *Tangle* due to the segmentation of that message in two transactions. Also, the likely range of variation (the interquartile range or IQR) is more compact for message $u$ for the same reason.

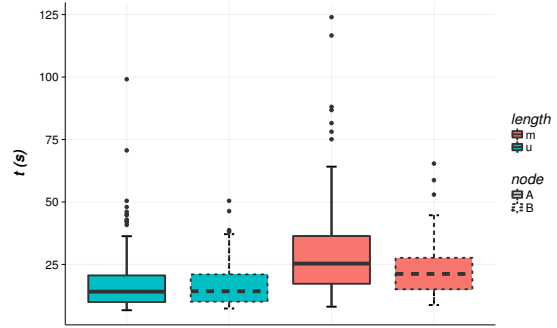

fig. 1: Time distribution for adding a message $u$ of size 1093 trytes and a message $m$ of size 2405 trytes using a public node $A$ and our private node $B$.

Regarding the variation in the delay as a function of the node to which the transaction is added, considering message $m$, node B exhibits a clear improved timing performance when compared to node A. This result can be justified by the comparatively reduced load in transaction requests that the private node B experiences. This gain is not so evident when performing transactions with $u$ messages.

Also represented on Fig. 1 are several transaction attachment delay outliers. These can have multiple causes, being the most likely one the occurrence of highly challenging "proof-of-work" cases.

To get a deeper understanding of the timings contributing to the overall delay in the transaction attachment to the *Tangle*, this procedure can be divided in three stages. First one needs to figure out which two transactions are going to be approved. This is done by running the tip selection algorithm two times. Secondly, for a transaction to be added to the Tangle, some "proof-of-work" must be carried out. The duration of this process has a high variance due to the randomness of finding a suitable nonce. Lastly, the transaction must be broadcasted to the network. The Cumulative Distribution Functions (CDF) of the latency experienced in these three phases are plotted for messages $u$ (1093 trytes) and $m$ (2405 trytes) in Fig. 2 and Fig. 3.

In both cases, what has an higher impact on the overall transaction insertion delay is the "attach to tangle" component corresponding to the "proof-of-work" that needs to be carried out in each transaction. The second most significant contribution to the overall delay is the "tip selection". The message broadcast averages around 1 second and, therefore, is negligible when compared to the other contributions.

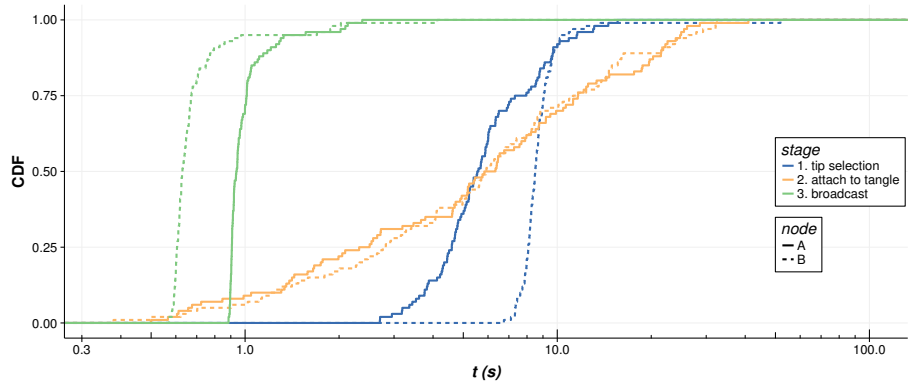

Fig. 2: CDF of the latency experienced during the multiple phases of adding a message $u$ of size 1093 trytes to the *Tangle* using the public node $A$ and our private node $B$.

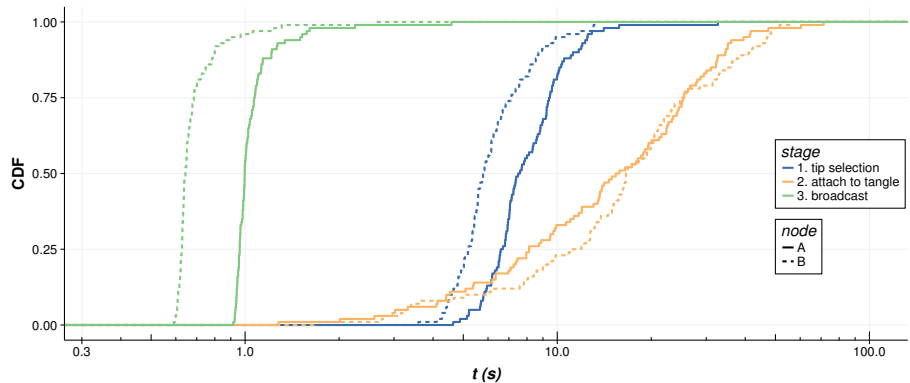

Fig. 3: CDF of the latency experienced during the multiple phases of adding a message $m$ of size 2405 trytes to the Tangle using the public node $A$ and our private node $B$.

A comparison between Fig. 2 and Fig. 3 reveals an expected behavior: the "proof-of-work" of $m$ messages exhibit a significant delay increase when compared $u$ messages. For example, considering the private node B, 25% of all message $m$ transactions experience a "attach to tangle" delay bellow 11 s, while the same is true for 75% all transactions encompassing message $u$.

The option for using the public node A or the private node B also affects the delays experienced in the different phases of the *Tangle* transaction attachment. As shown, with the exception of the "attach to tangle" phase, where the delays on both nodes have closer values, there is a noticeable delay variation regarding the "tip selection" and "broadcast" phases. In the former phase, for shorter messages $u$, this delay is generally smaller in the public node, while the opposite is true for larger messages $m$.

*2) Append MAMs to the Tangle:* As previously introduced, MAMs differ from regular transactions by being encrypted before their inclusion into the Tangle. In this sense, the end-to-end delay of attaching a MAM transaction and reading its content involves several stages, namely: "encoding", "tip selection", "attach to tangle", "broadcast" and "get message". The first and the last stage have not yet been introduced. The first, as the name suggests, corresponds to the encryption of the message via a private key. The last is the delay of reading a specific MAM.

Fig. 4 illustrates the latency CDF for each stage of the process of adding and fetching a MAM. As documented, the "tip selection" and "attach to tangle" phases are, as in regular transactions, the dominant contributors to the overall delay. The message encoding is conducted with a delay of around 380 milliseconds in average. Both the "broadcast" and "get message" stages have also negligible jitter and exhibit average delays of approximately 650 milliseconds and 950 milliseconds, respectively.

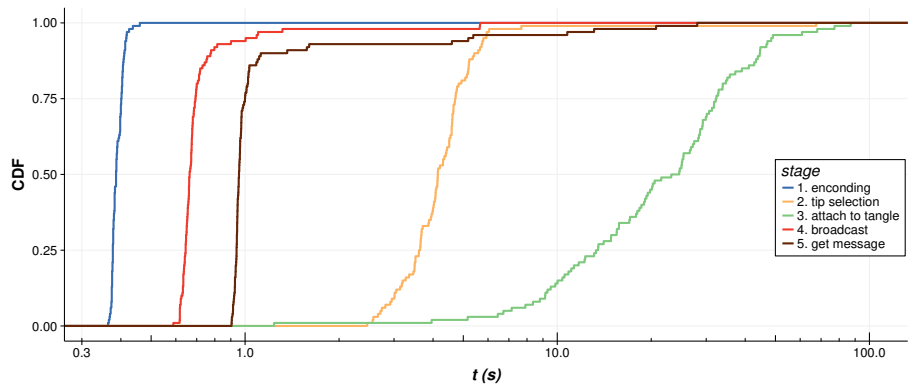

Fig. 4: CDF of the latency taken at each stage of the process of adding and fetching a message using the Masked Authenticating Messaging protocol.

*C. Discussion*

The above results show that the attachment of transactions to the tangle can be realized with a relatively low delay when compared to blockchains. The observed performance is set to enable vehicular applications with the benefits of the IOTA distributed ledger technology.

At the current "Beta" stage of the *Tangle* technology, the transactions addressed in this paper were not immediately *Confirmed* by the *Coordinator*. Indeed, many of them took a long time to be *Confirmed*. Despite the existence of mechanisms to promote a faster confirmation, the lack of direct or indirect approval by a "Milestone" can represent a security risk (e.g., double-spending) for any transaction. This risk can be circumvented by employing MAM messages among trusted parties.

One observation emerging from the results is that the overall delay of attaching transactions and MAMs to the *Tangle* is highly influenced by the "tip selection" algorithm and from the "proof-of-work" required in the process. The tip selection algorithm can be changed to become faster, although at the expense of inserting transactions in branches with less cumulative weigh and, therefore, with less probability of being validated by the network.

Regarding the "proof-of-work", it has been demonstrated in [21] that a delay of 300 milliseconds for this phase is feasible using a Raspberry Pi together with an FPGA. Such approach can be adopted to reduce the delay for this task by approximately one order of magnitude.

IV. CONCLUSIONS

This paper presented the *Tangle* as a possible solution to address the shortcomings of public blockchain technologies for vehicular applications. In this sense, the application context and a review of key vehicular research contributes was presented, as well as an introduction to the *Tangle* technology. The paper identified key operational performance parameters that were evaluated and discussed. The conclusion is that the *Tangle* exhibits smaller transaction delays than existing public blockchains. Another conclusion is that the performance of encrypted Masked Authenticated Messages exhibits a performance comparable with regular *Tangle* transactions. This will enable the support of privacy in vehicular communications with negligible latency overhead.

Future work will be focused on extending the presented analysis by realizing a larger set of trials (1000) per performance parameter in order to gain further insight and corroborate the observed results. Mechanisms for shortening the "proof-of-work" and tip selection delays will be researched with focus on embedded systems that can be easily adapted for vehicular applications.

REFERENCES

[1] Andy Greenberg, "Hackers Fool Tesla S's Autopilot to Hide and Spoof Obstacles", Wired, April 2016, https://www.wired.com/2016/08/hackers-fool-tesla-ss-autopilot-hide-spoof-obstacles/ [Online]. Accessed in 14/07/2018.

[2] Zeljka Zorz, "Researchers hack BMW cars, discover 14 vulnerabilities", HelpNetSecurity, May 2018, https://www.helpnetsecurity.com/2018/05/23/hack-bmw-cars/ [Online]. Accessed in 14/07/2018.

[3] Andy Greenberg, "Hackers Remotely Kill a Jeep on the Highway With ME IN IT", Wired, July 2015, https://www.wired.com/2015/07/hackers-remotely-kill-jeep-highway/ [Online]. Accessed in 14/07/2018.

[4] Liam Tung, "VW-Audi security: Multiple infotainment flaws could give attackers remote access", ZDNet, May 2018, https://www.zdnet.com/article/vw-audi-security-multiple-infotainment-flaws-could-give-attackers-remote-access/ [Online]. Accessed in 14/07/2018.

[5] P. C. Bartolomeu and J. Ferreira, "Blockchain Enabled Vehicular Communications: Fad or Future?", DEWCOM, Vehicular Technology Conference 2018. *To appear.*

[6] Institute of Electrical and Electronics Engineers, IEEE Standard for Wireless Access in Vehicular Environments Security Services for Applications and Management Messages, IEEE Std 1609.2-2016 (Revision of IEEE Std 1609.2-2013), pp. 1240, March 2016.

[7] The European Telecommunications Standards Institute, Intelligent Transport Systems (ITS); Security; Security header and certificate formats, ETSI TS 103 097 V1.3.1 (2017-10), 2017.

[8] B. Fernandes, J. Rufino, M. Alam, and J. Ferreira, Implementation and Analysis of IEEE and ETSI Security Standards for Vehicular Communications, Mobile Networks and Applications, Feb 2018. [Online]. Available: https://doi.org/10.1007/s11036-018-1019-x

[9] M. Khodaei, H. Jin and P. Papadimitratos, "SECMACE: Scalable and Robust Identity and Credential Management Infrastructure in Vehicular Communication Systems," in IEEE Transactions on Intelligent Transportation Systems, vol. 19, no. 5, pp. 1430-1444, May 2018.

[10] Puthal, N. Malik, S. P. Mohanty, E. Kougianos, and C. Yang, The Blockchain as a Decentralized Security Framework [Future Directions], IEEE Consumer Electronics Magazine, vol. 7, no. 2, pp. 1821, March 2018.

[11] Aste, P. Tasca, and T. D. Matteo, Blockchain Technologies: The Foreseeable Impact on Society and Industry, Computer, vol. 50, no. 9, pp. 1828, 2017.

[12] S. Nakamoto, "Bitcoin: A peer-to-peer electronic cash system", 2008 (Accessed June 10, 2018), bitcoin.org. [Online]. Available: https://bitcoin.org/bitcoin.pdf

[13] N. Lasla, M. Younis, W. Znaidi, and D. B. Arbia, Efficient Distributed Admission and Revocation Using Blockchain for Cooperative ITS, in 2018 9th IFIP International Conference on New Technologies, Mobility and Security (NTMS), Feb 2018, pp. 15.

[14] A. Dorri, M. Steger, S. S. Kanhere, and R. Jurdak, BlockChain: A Distributed Solution to Automotive Security and Privacy, IEEE Communications Magazine, vol. 55, no. 12, pp. 119125, DECEMBER 2017.

[15] A. Lei, H. Cruickshank, Y. Cao, P. Asuquo, C. P. A. Ogah, and Z. Sun, Blockchain-Based Dynamic Key Management for Heterogeneous Intelligent Transportation Systems, IEEE Internet of Things Journal, vol. 4, no. 6, pp. 18321843, Dec 2017.

[16] Z. Yang, K. Yang, L. Lei, K. Zheng, and V. C. M. Leung, Blockchain-based Decentralized Trust Management in Vehicular Networks, IEEE Internet of Things Journal, pp. 11, 2018.

[17] M. Singh and S. Kim, Trust Bit: Reward-Based Intelligent Vehicle Communication Using Blockchain, in 2018 IEEE 4th World Forum on Internet of Things (WF-IoT), Feb 2018, pp. 6267.

[18] Park, C. Sur, H. Kim, and K.-H. Rhee, A Reliable Incentive Scheme Using Bitcoin on Cooperative Vehicular Ad Hoc Networks, IT CoNvergence PRActice (INPRA), vol. 5, no. 4, pp. 3441, December 2017.

[19] L. Li, J. Liu, L. Cheng, S. Qiu, W. Wang, X. Zhang, and Z. Zhang, CreditCoin: A Privacy-Preserving Blockchain-Based Incentive Announcement Network for Communications of Smart Vehicles, IEEE Transactions on Intelligent Transportation Systems, pp. 117, 2018.

[20] S. Popov, "The *Tangle*", IOTA Whitepaper version 1.3, October 2017, http://iotatoken.com/IOTA_Whitepaper.pdf [Online]. Accessed in 11/07/2018.

[21] Thomas Pototschnig, "IOTA PoW Hardware Accelerator FPGA for Raspberry Pi (und USB)", April 2018, https://microengineer.eu/2018/04/25/iota-pearl-diver-fpga/ [Online]. Accessed in 14/07/2018.